\documentclass[sn-mathphys-num]{sn-jnl}

\usepackage{graphicx}%
\usepackage{subcaption}
\usepackage{multirow}%
\usepackage{amsmath,amssymb,amsfonts}%
\usepackage{amsthm}%
\usepackage{mathrsfs}%
\usepackage[title]{appendix}%
\usepackage{xcolor}%
\usepackage{textcomp}%
\usepackage{manyfoot}%
\usepackage{booktabs}%
\usepackage{algorithm}%
\usepackage{algorithmicx}%
\usepackage{algpseudocode}%
\usepackage{listings}%
\usepackage[T1]{fontenc}
\usepackage[utf8]{inputenc}
\usepackage{tikz}
\usepackage{tablefootnote}

\theoremstyle{thmstyleone}%
\theoremstyle{thmstyletwo}%

\theoremstyle{thmstylethree}%

\begin{document}

\title{%
  \begin{minipage}\linewidth
    \centering
    CGRclust: Chaos Game Representation\\ for Twin Contrastive  Clustering\\ of Unlabelled DNA Sequences
  \end{minipage}
}

\author*[1]{\fnm{Fatemeh} \sur{Alipour}}\email{falipour@uwaterloo.ca}



\author[2]{\fnm{Kathleen A.} \sur{Hill}}\email{khill22@uwo.ca}

\author[1]{\fnm{Lila} \sur{Kari}}\email{lila.kari@uwaterloo.ca}

\affil*[1]{\orgdiv{School of Computer Science}, \orgname{University of Waterloo}, \orgaddress{\country{Canada}}}


\affil[2]{\orgdiv{Department of Biology}, \orgname{University of Western Ontario}, \orgaddress{\country{Canada}}}




\abstract{\textbf{Background:}  
Traditional supervised learning methods applied to DNA sequence taxonomic classification rely on the labor-intensive and time-consuming step of labelling the primary DNA sequences. Additionally, standard DNA classification/clustering methods involve time-intensive  multiple sequence alignments, which impacts their applicability to large genomic datasets or distantly related organisms. These limitations indicate a need for robust, efficient, and scalable unsupervised DNA sequence clustering methods that do not depend on sequence labels or alignment.

\textbf{Results:} This study proposes CGRclust, a novel combination of unsupervised twin contrastive clustering of  Chaos Game Representations (CGR) of DNA sequences, with convolutional neural networks (CNNs). 
To the best of our knowledge, CGRclust is the first method to use unsupervised learning for image classification (herein applied to two-dimensional CGR images) for clustering datasets of DNA sequences. CGRclust overcomes the limitations of traditional sequence classification methods by leveraging unsupervised twin contrastive learning to detect distinctive sequence patterns, without requiring DNA sequence alignment or biological/taxonomic labels. 
CGRclust accurately clustered twenty-five diverse datasets,  with sequence lengths ranging from 664 bp to 100 kbp, including mitochondrial genomes of fish, fungi, and protists, as well as viral whole genome assemblies and synthetic DNA sequences. Compared with three recent  clustering methods for DNA sequences (DeLUCS, \textit{i}DeLUCS, and MeShClust v3.0.),  CGRclust is the only method that surpasses 81.70\% accuracy across all four taxonomic levels tested for mitochondrial DNA genomes of fish. Moreover, CGRclust also consistently demonstrates superior performance across all the viral genomic datasets. The high clustering accuracy of CGRclust on these twenty-five datasets, which vary significantly in terms of sequence length, number of genomes, number of clusters, and level of taxonomy, demonstrates its robustness, scalability, and versatility.

\textbf{Conclusion:} CGRclust is a novel, scalable, alignment-free DNA sequence clustering method that uses CGR images of DNA sequences and CNNs for twin contrastive clustering of unlabelled primary DNA sequences, achieving superior or comparable accuracy and performance over current approaches. CGRclust demonstrated enhanced reliability, by consistently achieving over 80\% accuracy in more than 90\% of the datasets analyzed. In particular, CGRclust performed especially well in clustering viral  DNA datasets, where it consistently outperformed all competing methods.}




\maketitle
\section{\textcolor{black}{Introduction}}\label{sec1}
DNA sequence classification is essential for genomic analyses, contributing to the identification of evolutionary relationships, functional elements, and genetic variants, through the detection of sequence similarity. Conventional methods for classifying DNA sequences typically depend on labor-intensive and expert-mediated labelling of primary DNA sequences to determine sequence origin, function, and type. Furthermore, the stability of genome labels can be questioned, as taxonomic labels are not always definitive due to the absence of a clear taxonomic ``ground truth''~\cite{Sambucus,taxonomy}. Moreover, most traditional DNA sequence classification and clustering methods are alignment-based. The time complexity of DNA sequence alignment~\cite{msa_complexity}, coupled with a dependence on additional sequence information such as sequence homology~\cite{alignment_free}, makes these methods unsuitable for analyzing large or evolutionarily divergent genomic datasets.  These challenges emphasize the importance of developing robust and flexible alignment-free unsupervised approaches to DNA sequence classification that do not rely on DNA sequence labels, annotation, or alignment.

\textcolor{black}{In 1990, Jeffery introduced Chaos Game Representation (CGR), a method for mapping one-dimensional DNA sequences into two-dimensional space using chaotic dynamics~\cite{CGR,fractals}.} \textcolor{black}{A CGR maps each DNA sequence to a unique image. The process begins with a unit square whose corners are labelled A, C, G, and T, in a clockwise order starting from the bottom-left corner. The initial point in any CGR plot is the center of this square. To generate the CGR for a specific DNA sequence, the sequence is read from left to right, one nucleotide at a time. For each nucleotide read, a point is plotted midway between the previous point and the corner labelled with that nucleotide.} Several studies~\cite{alignment_free,lochel,additive_methods} have demonstrated that CGRs can act as genomic signatures, defined by Karlin and Burge~\cite{genomic_signature} as numerical quantities that can distinguish closely from distantly related organisms based on DNA sequence identity. The distance   between CGRs of DNA sequences can be computed using various metrics, e.g., Euclidean distance,  and can then be used for alignment-free comparisons and phylogeny construction to demonstrate evolutionary relationships within a group of organisms. Due to these properties, CGR has been considered a milestone in graphical bioinformatics~\cite{milestones,ModMap}. 

\begin{figure}[h]
\centering
\includegraphics[width=\textwidth]{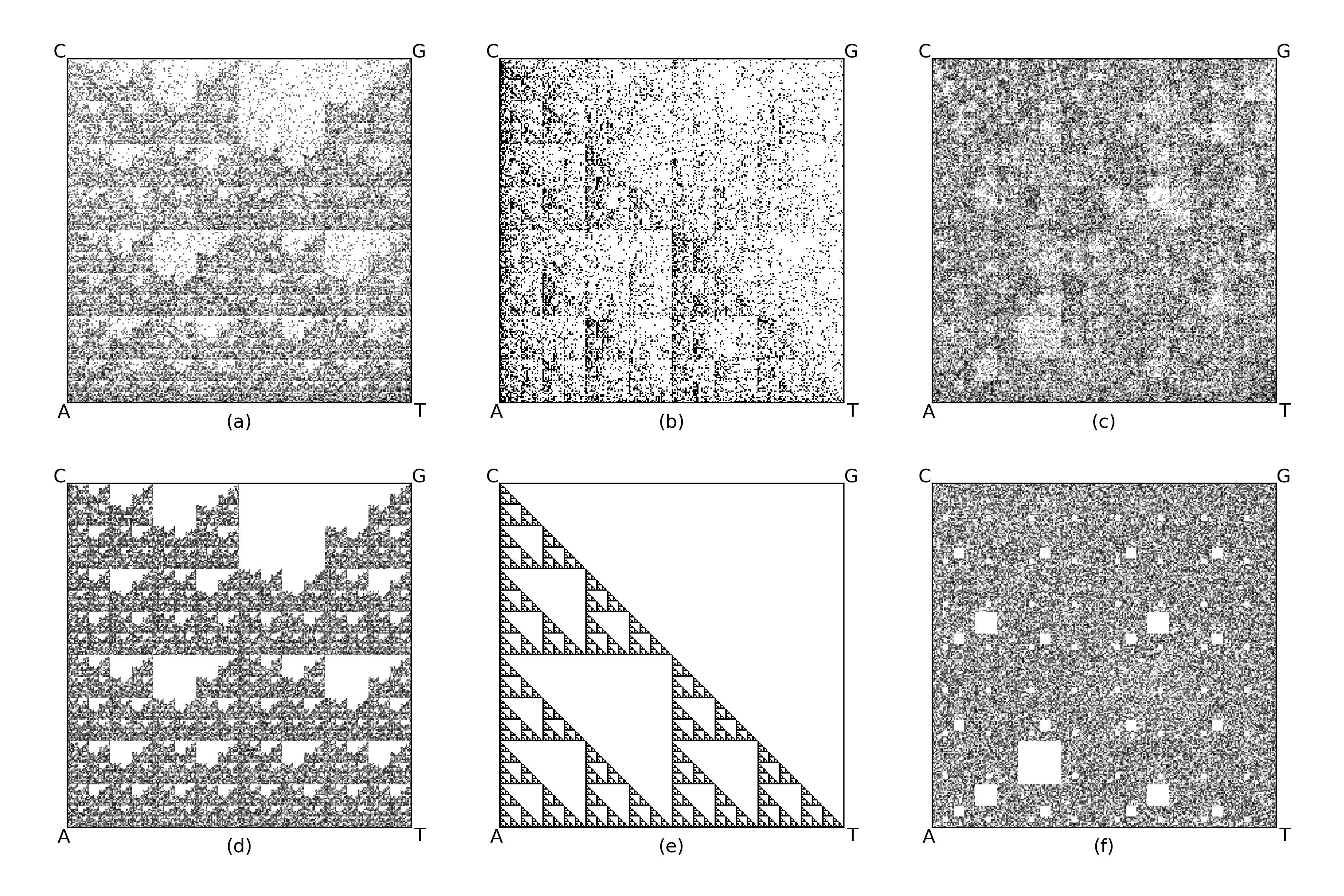}
\caption{Frequency Chaos Game Representation (FCGR) at resolution $k=8$ (for visualization purposes) of \textbf{(a)} human beta globin region on chromosome 11 of length 73,308 bp (Accession ID: U01317.1); \textbf{(b)} complete genome of \textit{Homo sapiens} isolate LI-T1 mitochondrion of length 16,566 bp (Accession ID: KX228192.1); \textbf{(c)} {\it Escherichia coli} plasmid of JE86-ST05 DNA with length 114,953 (Accession ID: AP022816.1);  Computer-generated  ``random" DNA sequences of length 100,000 bp avoiding substrings: \textbf{(d)}  ``CG'', \textbf{(e)}  ``G'', \textbf{(f)} ``CTA.'' } \label{fig:CGR-avoided-patterns}
\end{figure}

\textcolor{black}{Frequency CGR (FCGR), a quantified variant of CGR, divides the CGR into smaller squares to calculate and display the frequency of nucleotides within each segment. An FCGR at resolution \( k \) creates a \( 2^k \times 2^k \) numerical matrix which can be presented as a grayscale image wherein pixel intensities represent \( k \)-mer frequencies. Consequently, FCGR provides a compressed representation of DNA sequences and facilitates the analysis of distinct genomic signatures across different species.} \textcolor{black}{Figure~\ref{fig:CGR-avoided-patterns} illustrates some examples of FCGRs at resolution $k=8$ (selected for visualization purposes)  of real genomic DNA sequences, side by side with FCGRs of computer-generated DNA sequences.} \textcolor{black}{FCGR representations} of DNA sequences have been used in many alignment-free genome comparison applications, overcoming the quadratic runtime and scalability problems associated with alignment-based methods~\cite{alignment-free,lochel,Deschavanne,hill}. The use of FCGR permits alignment-free genomic sequence comparisons, when used in conjunction with digital signal processing techniques \cite{signal_processing,signal_processing_2} and machine learning methods \cite{rizzo,spiking_neural_networks,hpv,comparative_nucleosome,emam2020detection,covid}.

FCGR's ability to convert variable-length sequences into fixed-size dimensions is a key capability for machine learning, especially in DNA classification using convolutional neural networks (CNNs)~\cite{CNN}. \textcolor{black}{In a study by Rizzo et al.~\cite{rizzo}}, CNNs outperformed Support Vector Machines (SVMs) in classifying FCGR images of bacterial 16S gene sequences for both full-length sequences and 500 bp fragments. \textcolor{black}{Moreover, Safoury et al.~\cite{safoury} achieved an accuracy of 87\% with a simple CNN in classifying FCGRs of 660 DNA sequences across eleven genomic datasets.} In 2023, Avila et al. effectively classified SARS-CoV-2 DNA sequences into eleven clades using FCGR and CNNs~\cite{CouGaR-g}, achieving 96.29\% accuracy utilizing a ResNet50 neural network~\cite{resnet} and outperforming Covidex~\cite{covidex}, a random forest-based clade assignment tool. 
\textcolor{black}{Hammad et al.~\cite{hybrid} introduced} a hybrid CGR-based approach for detecting COVID-19, analyzing both whole and partial genome sequences of 7,951 human coronaviruses using AlexNet, Lasso algorithm, and KNN classifier. In spite of the effectiveness of these DNA classification methods, their reliance on labelled data is a significant limitation which highlights the urgent need for unsupervised algorithms that can perform well without the need of DNA sequence labels. 

To address this gap, dense neural networks have been used in conjunction with  FCGR for the  unsupervised clustering of DNA sequences in large, diverse datasets (up to 9,027 genomes) across different taxonomic levels and genetic distances~\cite{DeLUCS,iDeLUCS}. However,  in both DeLUCS~\cite{DeLUCS} and \textit{i}DeLUCS~\cite{iDeLUCS}, the neural networks first flattened each two-dimensional FCGR into a one-dimensional $k$-mer frequency vector. As a result, the features of two-dimensional FCGR images were not fully exploited in these methods. Another approach to clustering unlabelled DNA sequences,   MeShClust v3.0~\cite{meshclust}, used the mean-shift algorithm for generating pairwise identity scores without alignment. MeShClust v3.0 is built on its predecessors, MeShClust v1.0~\cite{meshclust1} (a DNA clustering method) and \textit{Identity}~\cite{identity} (a sequence alignment identity score predictor), and can efficiently cluster both long sequences, up to 3.7 million basepairs, and large datasets containing up to a million sequences. MeShClust v3.0 was tested on twenty-seven datasets, including twenty-two synthetic datasets and five real biological datasets, such as the human microbiome and maize transposons. 
\textcolor{black}{In spite of this progress, DeLUCS, \textit{i}DeLUCS, and MeShClust v3.0 underperform in clustering astrovirus sequences when compared to $K$-means++~\cite{astrovirus}}, even though they were previously validated on other viral datasets. These limitations highlight the need for the development of more robust approaches that can effectively manage the complexities of genetic diversity of a wide range of genomic datasets.

This paper presents CGRclust, a DNA sequence clustering method designed to identify discriminative features of DNA sequences, using two-dimensional FCGR images as the input to convolutional neural networks (CNNs), to fully leverage the information in this powerful DNA encoding. The clustering process in this study employs {\it twin contrastive learning} (TCL)~\cite{twin_clustering,contrastive_clustering}, a method proven effective in clustering images and text, which optimizes two contrastive learning objectives simultaneously—one at the instance-level and another at the cluster-level.


CGRclust's accuracy was evaluated across twenty-five datasets against DeLUCS~\cite{DeLUCS}, \textit{i}DeLUCS~\cite{iDeLUCS}, and MeShClust v3.0~\cite{meshclust}. Its clustering capabilities were tested on 2,688 mtDNA genomes of Cypriniformes, as well as five different viral genome datasets, including astroviruses, dengue virus, hepatitis C virus, and HIV-1. Furthermore, CGRclust was also assessed using mtDNA genomes from insects, protists, and fungi~\cite{iDeLUCS}, along with synthetic DNA sequences~\cite{meshclust}. All DNA sequences were unlabelled, with their  taxonomic labels used solely for post-hoc accuracy evaluation.

In summary, CGRclust is a novel, scalable, alignment-free clustering method that uses FCGR images and CNNs, for twin contrastive clustering of unlabelled primary DNA sequences. 
The main contributions of this paper are:
    \begin{itemize}
\item Being, to best of our knowledge, the  first application of twin contrastive learning to the clustering of DNA sequences, without requiring sequence homology, sequence labels, or sequence-length similarity.
\item Highly accurate clustering of a current dataset of 2,688 unlabelled fish mtDNA assemblies (order Cypriniformes). Clustering was performed at four different taxonomic levels, and CGRclust consistently achieved accuracy greater than 81.70\% at all levels.  This was either higher than, or comparable to, clustering accuracies of  the other state-of-the-art clustering methods (DeLUCS, \textit{i}DeLUCS, MeShClust v3.0).
\item Highly accurate clustering of several current datasets of unlabelled viral whole genomes (Astroviridae family into genera; dengue, HCV, HIV-1 species into virus subtypes), with accuracies ranging from 81.77\% to 100\% \textcolor{black}{(no classification error)}, surpassing the other state-of-the-art clustering methods.
\item Effective handling of challenging cases, such as unbalanced data, and scenarios with a high number of clusters and a small number of samples per cluster.
\item Superior or competitive accuracies compared to state-of-the-art methods on their benchmark datasets of unlabelled DNA sequences, e.g.,  73.56\% for insect mtDNA, 85.50\% for protist mtDNA, and 97.10\% for fungi mtDNA.
Furthermore, CGRclust consistently exceeded 92.26\% accuracy in clustering unlabelled synthetic DNA sequences of different lengths and identities.
    
\end{itemize}

\section{\textcolor{black}{Materials and }Methods}\label{sec2}
This section starts with a description of the datasets utilized in this study. This is followed by an overview of the proposed computational pipeline for contrastive clustering of DNA sequences in CGRclust. Chaos Game Representation (CGR), the graphical representation of DNA sequences used in this paper, is then defined, together with its quantified variant FCGR. Next, a description of the data augmentation strategies  used for this graphical representation (generation of {\it mimic sequences}) is presented, serving as the initial component of CGRclust's pipeline. Afterwards, the core concept of twin contrastive learning, details about the backbone model, and the majority voting scheme adapted to clustering FCGRs of DNA sequences are described. Lastly, details of implementation and testing are provided.

\subsection{Datasets}

To comprehensively evaluate the performance of CGRclust in clustering DNA sequences, we strategically selected four groups of datasets, comprising diverse genomic data both real and synthetic. The selection rationale was driven by the need to assess the clustering method across different levels of taxonomy with different degrees of relatedness, genomic conservation, and evolutionary dynamics. The Group 1 dataset includes mitochondrial DNA of fish, while the Group 2 dataset includes viral whole genomes.  Additionally, to facilitate direct comparisons with established methodologies, we incorporated datasets previously analyzed \textcolor{black}{by Mill{\'a}n et al}.~\cite{iDeLUCS} and \textcolor{black}{by Girgis}~\cite{meshclust} (Group 3 and Group 4 datasets, respectively). \textcolor{black}{In the following, the test labels are integrally linked with datasets and are used for ease of reference when discussing results.}

 The Group 1 dataset comprised complete mitochondrial DNA (mtDNA) sequences of Cypriniformes (an order of ray-finned fish).  This dataset was retrieved from the National Center for Biotechnology Information (NCBI) on January 30, 2024, with a filter selecting mtDNA sequences of length between 4 kbp and 25 kbp. Following the removal of ``partial'' and ``unverified'' genomes, 
 2,688 complete mitochondrial genomes of Cypriniformes were collected. At each taxonomic level, the cluster with the highest number of sequences was selected for the lower taxonomic level clustering task.  Due to significant variability and imbalance in the number of available sequences across the four taxonomic levels, sequences from clusters with fewer than 50 sequences were discarded. To address the imbalance, in the first three computational tests (Tests 1-3), we established a threshold based on the minimum number of sequences available in a cluster and randomly selected an equivalent number of sequences from the other clusters. Balancing the clusters was not needed in Test 4, as the dataset was already evenly distributed. Table~\ref{tab:merged-datasets} summarizes the dataset details for the Group 1 dataset (Cypriniformes mtDNA). The selection of this group of datasets was motivated by the conservative nature of mtDNA, which is predominantly coding and thus provides a stable framework for assessing clustering methodologies at multiple taxonomic levels. The uniformity of high conservation over the mtDNA genome compared to the regional variation in sequence conservation of the nuclear genome, coupled with its wide use in phylogenetic studies~\cite{mtDNA,mtDNA2}, makes mtDNA data an ideal candidate for initial clustering evaluations. 

\begin{table}[htp]
\caption{\textcolor{black}{\textbf{Details of the Group 1, 2, and 3 dataset in Tests 1 through 13.}} Underlined font indicates the cluster with the highest number of sequences, which was subsequently selected for clustering at a lower taxonomic level.} \label{tab:merged-datasets}
\centering
\begin{tabular}{|c|l|c|c|c|c|}
\hline
\multicolumn{6}{|c|}{\textbf{Group 1: Cypriniformes Full Mitochondrial Genomes}} \\
\hline
Test & \multicolumn{1}{c|}{\begin{tabular}[c]{@{}c@{}}{\bf Taxonomic Clustering}\\ (No. of seq. per cluster)\end{tabular}} & \begin{tabular}[c]{@{}c@{}}No. of \\ seq.\end{tabular} & \begin{tabular}[c]{@{}c@{}}Min. seq.\\ len. (bp)\end{tabular} & \begin{tabular}[c]{@{}c@{}}Avg. seq.\\ len. (bp)\end{tabular} & \begin{tabular}[c]{@{}c@{}}Max. seq.\\ len. (bp)\end{tabular} \\
\hline
1    & \begin{tabular}[t]{@{}l@{}}{\bf Order into Suborder}: Cypriniformes\\ into Catostomoidei, Cobitoidei,\\ \underline{Cyprinoidei} (166 each)\end{tabular} & 498 & 15,655 & 16,610 & 17,859 \\
2    & \begin{tabular}[t]{@{}l@{}}{\bf Suborder into Family}: Cyprinoidei\\ into Acheilognathidae, \underline{Cyprinidae},\\ Danionidae, Gobionidae, Leuciscidae, \\Xenocyprididae (105 each)\end{tabular} & 630 & 15,616 & 16,620 & 18,220 \\
3    & \begin{tabular}[t]{@{}l@{}}{\bf Family into Subfamily}: Cyprinidae\\ into Acrossocheilinae, \underline{Cyprininae},\\ Labeoninae, Poropuntiinae,\\ Schizopygopsinae, Schizothoracinae,\\ Smiliogastrinae, Torinae (56 each)\end{tabular} & 448 & 15,609 & 16,603 & 17,426 \\
4    & \begin{tabular}[t]{@{}l@{}}{\bf Subfamily into Genus}: Cyprininae\\ into Carassioides (74), Cyprinus (77),\\ Sinocyclocheilus (62) \end{tabular} & 213 & 16,562 & 16,592 & 17,426 \\
\hline
\multicolumn{6}{|c|}{\textbf{Group 2: Viral Whole Genomes}} \\
\hline
5    & \begin{tabular}[t]{@{}l@{}} {\bf Family into Genus [unbalanced]}: \\
Astroviridae into Avastrovirus (363),\\ Mamastrovirus (726)\end{tabular} & 1,089 & 5,003 & 6,653 & 8,324 \\
6    & \begin{tabular}[t]{@{}l@{}} {\bf Family to Genus [balanced]}:\\
Astroviridae into Avastrovirus,\\
Mamastrovirus (363 each)\end{tabular} & 726 & 5,003 & 6,787 & 7,960 \\
7    & \begin{tabular}[t]{@{}l@{}}{\bf Species into Subtypes}: dengue virus\\ 
into subtypes 1, 2, 3, 4 (407 each) \end{tabular} & 1,628 & 10,161 & 10,563 & 10,940 \\
8    & \begin{tabular}[t]{@{}l@{}}{\bf Species into Subtypes}: hepatitis C\\
virus into subtypes 1, 1a, 1b, 2b, 3a\\
(190 each) \end{tabular} & 950 & 7,005 & 9,059 & 9,678 \\
9    & \begin{tabular}[t]{@{}l@{}}{\bf Species into Subtypes}: human\\ 
immunodeficiency virus type 1 into\\
subtypes 01B, 01\_AE , 02\_AG, A1, A1C,\\
A1CD, A1D, A6, B, BF1, C, D, G \\(100 each) \end{tabular} & 1,300 & 8,001 & 8,904 & 9,839 \\
\hline
\multicolumn{6}{|c|}{\textbf{Group 3: mtDNA of Insects, Protists, and Fungi (from~\cite{iDeLUCS})}} \\
\hline
10    & \begin{tabular}[t]{@{}l@{}} {\bf Class into Order}: Insecta into\\
Lepidoptera, Hemiptera, Diptera,\\
Coleoptera, Dictyoptera, Orthoptera,\\
Hymenoptera (650 each)\end{tabular} & 4,550  & 14,602 & 15,897 & 25,011 \\
11    & \begin{tabular}[t]{@{}l@{}} {\bf Kingdom into Phylum}: \\Chromista/Plantae (Protista) into\\
Alveolata, Stramenopiles, Rhodophyta\\ (315 each)\end{tabular} & 945 & 5,498 & 24,697 & 24,697 \\
12    & \begin{tabular}[t]{@{}l@{}}{\bf Kingdom into Phylum}: Fungi into\\
Ascomycota, Basidiomycota (335 each) \end{tabular} & 670 & 22,528 & 59,864 & 99,976 \\
13    & \begin{tabular}[t]{@{}l@{}}{\bf Phylum into Subphylum}: Ascomycota\\ into Pezizomycotina,\\
Saccharomycotina (535 each)\end{tabular} & 1,070 & 20,063 & 63,388 & 99,850 \\
\hline
\end{tabular}
\end{table}

To further demonstrate the effectiveness of CGRclust, we assessed its performance across five viral whole genome datasets in the Group 2 dataset: an updated version of the virus family Astroviridae genomes~\cite{astrovirus} (Test 5) and its balanced version (Test 6), an updated version of whole genomes of dengue virus (Test 7), hepatitis C virus (HCV) (Test 8), and human immunodeficiency virus 1 (HIV-1) (Test 9) previously classified \textcolor{black}{by Solis-Reyes et al.}~\cite{Kameris} with supervised machine learning methods. Table~\ref{tab:merged-datasets} outlines the details of the Group 2 dataset. In Test 5, 1,089 complete astrovirus genomes were collected, for taxonomic clustering of the sequences from family to genus level. Test 6 uses a cluster-balanced variant of the astrovirus dataset to address the initial label imbalance, thereby ensuring that the clustering results are not skewed by this disparity. All astrovirus sequences were downloaded from NCBI on April 4, 2024, with a filter selecting genome lengths ranging between 5 kbp and 10 kbp. Furthermore, we addressed the clustering of viral sequences at a lower level of species to subtypes in Tests 7-9. This categorizing which is called viral subtyping is crucial for understanding intraspecific variation, tracking epidemiological trends, and developing targeted treatments or vaccines. The dengue virus sequences used in Test 7 were obtained from \url{https://www.ncbi.nlm.nih.gov/genomes/VirusVariation/Database/nph-select.cgi?taxid=12637} using the query parameters ``Nucleotide'', ``Full-length sequences only'', and ``Collapse identical sequences'', resulting in a dataset of 5,868  sequences. Following cluster balancing, we obtained a dengue dataset comprising 1,628 dengue virus whole genomes spanning four distinct subtypes. The HCV genomes utilized in Test 8 were sourced from the LANL sequence database, accessible at \url{https://hcv.lanl.gov/components/sequence/HCV/search/searchi.html}, with the query settings ``Excluding recombinants'', ``Excluding ‘no genotype‘'', ``Genomic region: complete genome'', and ``Excluding problematic'', resulting in 3,612 whole HCV genomes. After removing clusters with less than 100 sequences and balancing the dataset, we obtained 950 full HCV genomes spanning five different subtypes. Finally, the HIV-1 genomes in Test 9 were retrieved from the Los Alamos (LANL) sequence database, accessible at \url{https://www.hiv.lanl.gov/components/sequence/HIV/search/search.html} with query parameters  ``virus: HIV-1, genomic region: complete genome, excluding problematic,'' which resulted in a dataset comprising  20,525 HIV-1 full genomes. We then removed HIV-1 subtypes with fewer than 100 sequences and balanced the remaining subtypes, thus obtaining a dataset comprising 13,000 HIV-1 whole genome sequences spanning 13 subtypes. The three viral datasets used in Tests 7-9 were downloaded on April 1, 2024. Viral genomes are characterized by higher mutation rates and greater evolutionary diversity compared to the mtDNA, presenting distinct challenges for clustering algorithms. This variability tests the robustness and adaptability of CGRclust under conditions of rapid genomic changes and diverse evolutionary pressures.

Next, we evaluated the performance of CGRclust on three core datasets used \textcolor{black}{by Mill{\'a}n et al.}~\cite{iDeLUCS} (Group 3 dataset: mtDNA of Insects, Protists, and Fungi), as well as 12 synthetic DNA datasets analyzed \textcolor{black}{by Girgis}~\cite{meshclust} (Group 4 dataset: synthetic sequences). Including these datasets allowed for direct comparisons with existing studies, providing benchmarks against established clustering methods. The Group 3 dataset is described in Table~\ref{tab:merged-datasets}. Note that, given the observed mixed taxonomic levels used \textcolor{black}{by Mill{\'a}n et al.}~\cite{iDeLUCS}  for clustering the Fungi dataset, and the fact that both subphyla ``Pezizomycotina'' and ``Saccharomycotina'' belong to phylum Ascomycota, we divided this clustering task into two parts, Tests 12 and 13. The first task (Test 12) involved clustering kingdom Fungi into phyla ``Ascomycota'' and ``Basidiomycota'', while the second task (Test 13) focused on clustering phylum Ascomycota into subphyla ``Pezizomycotina'' and ``Saccharomycotina''. Details about the Group 4 dataset~\cite{meshclust} are presented in Table~\ref{tab:dataset-meshclust}. The sequence lengths of six datasets, each beginning with the prefix ``Medium-'' range between 653 and 2,062 bp, while the other six datasets, prefixed with ``Long-'', span from 1,393 to 4,049 bp. The numerical values ranging from 60 to 97 in the dataset labels represent the identity score, a measure of designed relatedness determined by the ratio of identical nucleotides in two sequences relative to the alignment length (including gaps). 
These synthetic sequences, designed with different sequence lengths and identity score thresholds, evaluate the performance of CGRclust under controlled, and different conditions. For further details on Group 3 and Group 4 datasets, the reader is referred to~\cite{iDeLUCS} and~\cite{meshclust}, respectively.

\begin{table}[h]
\setlength\tabcolsep{1pt}
\caption{\textbf{Details of the Group 4 dataset in Tests 14 through 25 (synthetic sequences with different length and identity score thresholds from~\cite{meshclust}).} The prefixes ``Medium-'' and ``Long-'' in the dataset names denote the length of the sequences they contain and
and the numerical values ranging from 60 to 97 in these names represent the identity score, indicating the percentage of similarity between the sequences.}\label{tab:dataset-meshclust}
\centering
\begin{tabular}{|c|c|c|c|c|c|c|c|c|c|}
\hline
Test &  \multicolumn{1}{|c|}{Dataset} & \begin{tabular}[t]{@{}c@{}}No. of \\ seq.\end{tabular} &  \begin{tabular}[t]{@{}c@{}}Min. seq.\\ len. (bp)\end{tabular} & \begin{tabular}[t]{@{}c@{}}Avg. seq.\\ len. (bp)\end{tabular} & \begin{tabular}[t]{@{}c@{}}Max. seq.\\ len. (bp)\end{tabular} & \begin{tabular}[t]{@{}c@{}}No. of \\ clusters\end{tabular} & \begin{tabular}[t]{@{}c@{}}Min. cluster\\ size\end{tabular} & \begin{tabular}[t]{@{}c@{}}Avg. cluster\\ size\end{tabular} & \begin{tabular}[t]{@{}c@{}}Max. cluster \\size\end{tabular} \\
\hline
14    & Medium-60 & 18,210 & 653 & 1,365 & 2,062 & 100 & 13 & 202 & 398 \\
\hline
15    & Medium-70 & 18,731  & 678 & 1,359 & 2,027 & 100 & 8 & 212 & 398\\
\hline
16    & Medium-80 & 20,939  & 664 & 1,425 & 2,043 & 100 & 14 & 222 & 398\\
\hline
17    & Medium-90 & 21,266  & 730 & 1,340 & 2,016 & 100 & 5 & 194 & 400\\
\hline
18    & Medium-95 & 24,039  & 724 & 1,446 & 2,038 & 100 & 7 & 203 & 396\\
\hline
19    & Medium-97 & 20,772  & 736 & 1,358 & 2,022 & 100 & 13 & 192 & 390\\
\hline
20    & Long-60 & 20,885  & 1,393 & 2,758 & 4,039 & 100 & 7 & 207 & 398\\
\hline
21    & Long-70 & 18,558  & 1,441 & 2,754 & 4,062 & 100 & 19 & 224 & 399\\
\hline
22    & Long-80 & 20,525  & 1,396 & 2,639 & 3,974 & 100 & 5 & 194 & 398\\
\hline
23    & Long-90 & 22,518  & 1,489 & 2,586 & 3,964 & 100 & 5 & 196 & 400\\
\hline
24    & Long-95 & 20,222  & 1,461 & 2,890 & 4,049 & 100 & 10 & 206 & 400\\
\hline
25    & Long-97 & 19,960  & 1,486 & 2,715 & 3,988 & 100 & 5 & 210 & 398\\
\hline
\end{tabular}
\end{table}


\subsection{Method overview}
The contrastive clustering method proposed in this paper, CGRclust, utilizes a quantified variant of CGR, a graphical encoding of DNA sequences introduced by Jeffrey~\cite{CGR}. This quantified DNA encoding, referred to as $FCGR$, represents a DNA sequence at resolution $k$ as a two-dimensional unit square image. In an FCGR, the intensity of each pixel signifies the frequency of a particular $k$-mer in the input DNA sequence~\cite{Deschavanne}. \textcolor{black}{For a formal definition of CGR and FCGR  see Supplementary Material 1.} To capture the positional information (location of points) within FCGR images, a CNN model was integrated into the pipeline. CGRclust enhances the clustering performance by leveraging unsupervised contrastive learning. Contrastive learning is a powerful technique that can learn informative representations by comparing how similar or different pairs of examples are, rather than relying solely on raw data or labelled examples~\cite{self-supervised}. This approach helps the 
model understand the underlying structures of the data by pulling similar instances (elements of a so-called ``positive pair") closer, while pushing dissimilar ones (elements of a  so-called ``negative pair")  farther apart in the representation space. Here, a {\it positive pair}  is defined as consisting of two ``augmented"  versions of an input DNA sequence, called {\it mimic sequences}. Mimic sequences are generated by the algorithm from an original DNA sequence so as to be similar to the original, or related to it in a meaningful way. In this context, a {\it negative pair} is defined as any other pair of sequences in the dataset. 
The clustering process in this study takes advantage of the concept of {\it twin contrastive learning} (TCL) ~\cite{twin_clustering,contrastive_clustering}, a method that simultaneously optimizes two contrastive learning objectives, one at the instance-level and another at the cluster-level, as detailed below.

\begin{figure}[h]
\includegraphics[width=\textwidth]{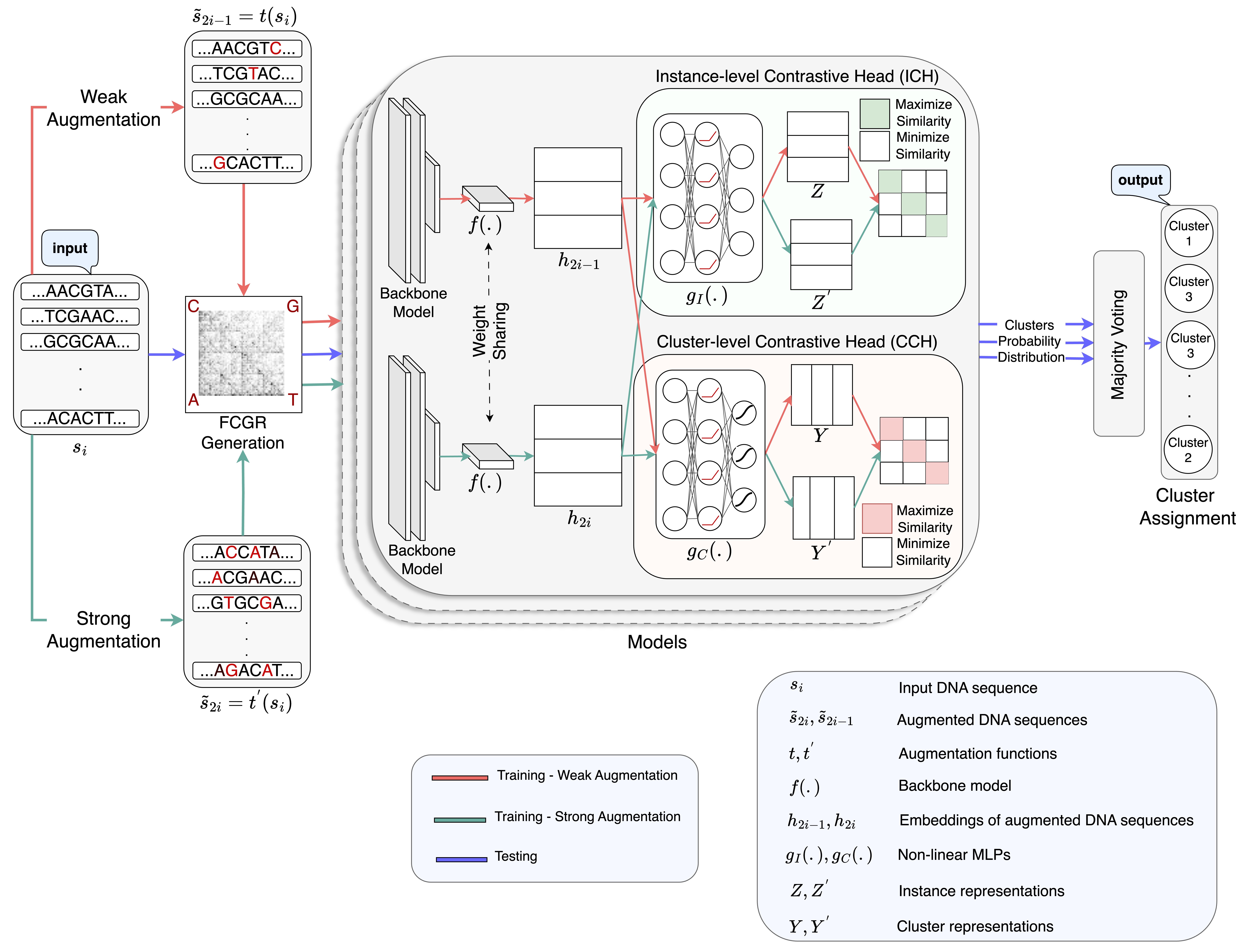}
\caption{\textbf{CGRclust pipeline:} Left panel: The process begins with data augmentation to create positive pairs (pairs of {\it mimic sequences}, pipeline component 1), followed by the generation of FCGR images of these augmented DNA sequences. Middle Panel:  The FCGR images are fed into the backbone model (CNN) for embedding into a latent feature space (pipeline component 2). The twin contrastive learning scheme employs an instance-level contrastive head (ICH) and a cluster-level contrastive head (CCH) to perform contrastive learning at both the instance and the cluster levels (pipeline component 3 and 4, respectively). 
Right panel: To counteract the inherent variance in CNN training outcomes, a majority voting strategy is applied, aggregating results from multiple CNN models with distinct initializations to finalize cluster assignments for each input DNA sequence.} \label{fig:pipeline}
\end{figure}

Figure~\ref{fig:pipeline} illustrates an overview of the proposed CGRclust pipeline. The pipeline consists of four main components: 1) data augmentation (generation of  mimic sequences) for FCGR positive pair construction, 2) backbone model for projection into a latent feature space , 3) {\it instance-level} contrastive head (ICH), and 4) {\it cluster-level} contrastive head (CCH). The first component is shown in the left panel of Figure~\ref{fig:pipeline}, while the other three components are in the middle panel. Initially, pairs of mimic sequences constructed during the data augmentation phase (pipeline component 1), and assumed to belong to the same cluster, are projected into a latent feature space using CNNs (pipeline component 2).  It is important to note that in the training phase, the two mimic sequences constructed from each original sequence were used as members of a positive pair, while the original sequence was used exclusively in the testing phase. Subsequently, the ICH (pipeline component 3) and CCH (pipeline component 4) conduct instance-level and cluster-level contrastive learning. ICH is designed to enhance the similarity of representations of positive pairs in the latent feature space, while making the representations of negative pairs more distinct. On the other hand, CCH's goal is to effectively separate clusters of data points, ensuring that each cluster is distinctly different from the others.

 The two components (ICH and CCH) are simultaneously optimized through twin contrastive learning (TCL) by operating on the row (ICH) and column  (CCH) spaces of the feature matrix, respectively. Through this simultaneous optimization, CGRclust enhances the representation's quality by handling both detailed (in ICH) and broad (in CCH) distinctions in the data, all without relying on pre-defined taxonomic labels. As the training process involves randomized algorithms leading to high variance outcomes depending on  the different initializations and random seeds,
a majority voting scheme is then employed (right panel of Figure~\ref{fig:pipeline}), which uses the outcomes of five distinct CNN models with different initializations to determine the final cluster assignment for each sequence.
 
To evaluate the quality of the clusters, an additional step, independent from the previous components, is conducted. This step utilizes the Hungarian algorithm~\cite{Hungarian}, \textcolor{black}{ a method that effectively pairs elements from two sets to minimize the overall mismatch}, to determine the optimal correspondence between the cluster assignments learned by the CGRclust and the actual taxonomic cluster labels. Subsequently, it evaluates the accuracy of the CGRClust predictions.

\subsection{DNA data augmentation: Mimic sequences}
 Data augmentation plays a critical role in contrastive clustering by significantly enhancing the model's ability to learn invariant representations from limited data. By adding different types of changes to the training data  (thereby generating positive pairs), data augmentation helps the model to focus on the key features that define each cluster, avoiding the trap of fitting too closely to random noise or unimportant details. Consequently, CGRclust is based on constructing positive pairs and negative pairs through data augmentations. A pair of positive data points is a pair of {\it mimic sequences},  that are considered to be similar or related in some meaningful way (e.g. belonging to the same cluster), while a pair of negative data points is a pair of sequences that are considered to be dissimilar. We adapted a similar approach to~\cite{twin_clustering}, and used an effective augmentation strategy by mixing weak and strong transformations as it previously showed superior performance on both image and text data when combined with TCL. For each DNA sequence input \( s_i \), we define transformations \( t \) and \( t' \) as follows: \( t \) and \( t' \) are functions from the domain of DNA sequences to the set of augmented DNA sequences, with \( t \) applying a set of transformations from an augmentation family \( T \),  and \( t' \) applying a set of transformations from an augmentation family  \( T' \).  These transformations are designed to modify the input sequence \( s_i \) in distinct ways, generating a positive pair represented as $(\tilde{s}_{2i-1},\tilde{s}_{2i})$, where $\tilde{s}_{2i-1} = t{(s_i)}$ and $\tilde{s}_{2i} = t^{'}(s_i)$.

 Note that direct image transformations traditionally used in computer vision for data augmentation (image flipping, cropping,  or rotation), if applied to CGR/FCGR images,  do not correspond to biologically meaningful or minor changes in the original DNA sequence. Indeed, such transformations could result in drastic and non-intuitive sequence changes, since the CGR/FCGR representations depend on the sequence's nucleotide order and composition. Thus,  in CGRclust we opted to modify raw DNA sequences to create {\it mimic sequences}.  This approach ensures that any resulting image alterations are meaningful, and mirror potential natural genetic variations in sequence composition.

 In CGRclust pipeline, data augmentations  were  implemented through functions  \( t \)  and \( t' \), belonging to the augmentation families \( T \) (weak augmentations)  and \( T' \) (strong augmentations) respectively.
  Two types of data augmentation were explored,  {\it mutation} and {\it fragmentation}.   Both mutation and fragmentation of a DNA sequence, when appropriately applied, can alter the sequence while still maintaining patterns within its FCGR that are very similar (but not identical) to the FCGR of the original DNA sequence. 
 
 Mutation, denoted by $mutate(\mu)$ has a mutation rate $\mu$ as parameter, and performs two types of substitution mutations (transitions and transversions) on the original DNA sequence. The probability of transitions  is defined as being $\mu$ while the probability of transversions is $0.5 * \mu$, as the mutational hypothesis holds that the transition mutation rates are higher than the transversion rates in practice~\cite{mutation}. Fragmentation, denoted by $frag(len)$, has the length $len$  of the desired fragment as parameter. Given a DNA sequence of length $n$ as input, fragmentation outputs a random fragment of length $len$ of the input sequence ($len \leq n$).

In each computational experiment, the augmentation functions $t$ and $t'$ can be either  mutation or a fragmentation. If the selected augmentation function is mutation, then $t$ is  the function $mutate(\mu_1)$  (weak), and $t'$ is the function  $mutate(\mu_2)$ (strong), where  $\mu_1 <\mu_2$. Similarly, if the selected augmentation function is fragmentation, then  the function \( t \) is  $frag(len_1)$ (weak), while  \( t' \) is the function  $frag(len_2)$ (strong), where  $len_2 < len_1$.

  To evaluate the impact of different data augmentation strategies on CGRclust,  both mutation and fragmentation were explored, each with   different values for their respective parameters.  Details on these computational experiments  can be found in Supplementary Material 2.  The final findings suggest that mutation outperforms fragmentation as a data augmentation function, and its optimal parameters were empirically determined to be  $\mu_1 = 10^{-4}$ for the weak augmentation, and $\mu_2 = 10^{-2}$ for the strong augmentation. Thus, mutation with these parameters was used as the default data augmentation and parameters for all computational experiments in this study.

 Given the constructed pairs, a shared backbone $f(\cdot)$ is used to extract features $h$ from the augmented samples (mimic sequences) through $h_{2i-1} = f(X_{\tilde{s}_{2i-1}})$ and $h_{2i} = f(X_{\tilde{s}_{2i}})$. To extract the important features of FCGR images, the backbone model was used to convert the two-dimensional input FCGRs into one-dimensional embeddings. Details about the backbone model used to process FCGR can be found in section 2.6 (Backbone model architecture).

\subsection{Twin contrastive learning (TCL)}
Inspired by~\cite{twin_clustering,contrastive_clustering}, during the training phase, the backbone, ICH, and CCH undergo joint optimization based on the following twin contrastive loss function:
\begin{equation}
    L_{train} = \alpha L_{ins} + (1 - \alpha) L_{clu}
\end{equation}
Here, $L_{ins}$ denotes the instance-level contrastive loss computed via ICH, to increase the similarity between positive pairs and decrease it between negative pairs. Meanwhile, $L_{clu}$ represents the cluster-level contrastive loss, determined through CCH, focusing on refining the pairwise similarities of cluster representations between weak and strong data augmentations. $\alpha$ represents a weighting parameter that balances the contributions of the instance-level contrastive loss ($L_{ins}$) and the cluster-level contrastive loss ($L_{clu}$) in the overall training loss ($L_{train}$).

The parameter $\alpha$ controls the relative importance of the two components during optimization. To determine its optimal value, we tested different values for this hyperparameter and it was empirically determined that the value of 0.7 for \(\alpha\) consistently delivered either the highest or close to the highest accuracy. Furthermore, it was observed that values within the range of 0.5 to 0.8 generally yielded superior outcomes, suggesting a robust zone of performance for \(\alpha\) across different data conditions. For additional details, the reader is referred to  Supplementary Material 2.


Optimal clustering would classify instance pairs within the same class as positive and those across classes as negative. Yet, in the absence of predefined labels, we adapt by forming mimic sequence instance pairs via data augmentations. Given a batch size of $N$, we subject each DNA sequence, $s_i$, to two variants of data augmentations, generating 2N augmented samples expressed as ${\tilde{s}_1, \tilde{s}_2,...,\tilde{s}_{2i-1}, \tilde{s}_{2i},...\tilde{s}_{2N}}$. Before employing ICH and CCH, we map features into two different subspaces using two-layer nonlinear Multilayer Perceptrons (MLPs), symbolized as $g_I(\cdot)$ and $g_C(\cdot)$, respectively. 

The InfoNCE loss~\cite{InfoNCE}, which includes a computational parameter so-called ``temperature parameter'' (\(\tau\)) to scale the contrastive loss, is applied to fine-tune both contrastive mechanisms. A comprehensive hyperparameter optimization of the twin deep clustering model focused on the instance- and cluster-level temperature parameters (\(\tau_I\) and \(\tau_C\)) within the ICH and CCH was conducted. Examining different values for each temperature parameter in the range $[0.1, 1]$, it was empirically determined that \(\tau_I=0.1\) and \(\tau_C=1.0\) consistently yield relatively high accuracy across all datasets. This advancement aligns with the hypothesis that a lower \(\tau_I\) encourages individual instance differentiation, aligning with the ICH's goal, while a higher \(\tau_C\) enhances group discrimination, mirroring the CCH's objective~\cite{temperature}.


While a confidence-based boosting strategy, which involves iterative adjustments to the learning process based on model prediction confidence, yielded a slight enhancement in the clustering outcomes of~\cite{twin_clustering}, no significant improvement was observed for FCGR clustering. Therefore, we opted against incorporating this step to maintain pipeline simplicity and efficiency. For additional information about TCL see Supplementary Material 3 and~\cite{twin_clustering}.

\subsection{Backbone model architecture}
\begin{figure}[h]
\includegraphics[width=\textwidth]{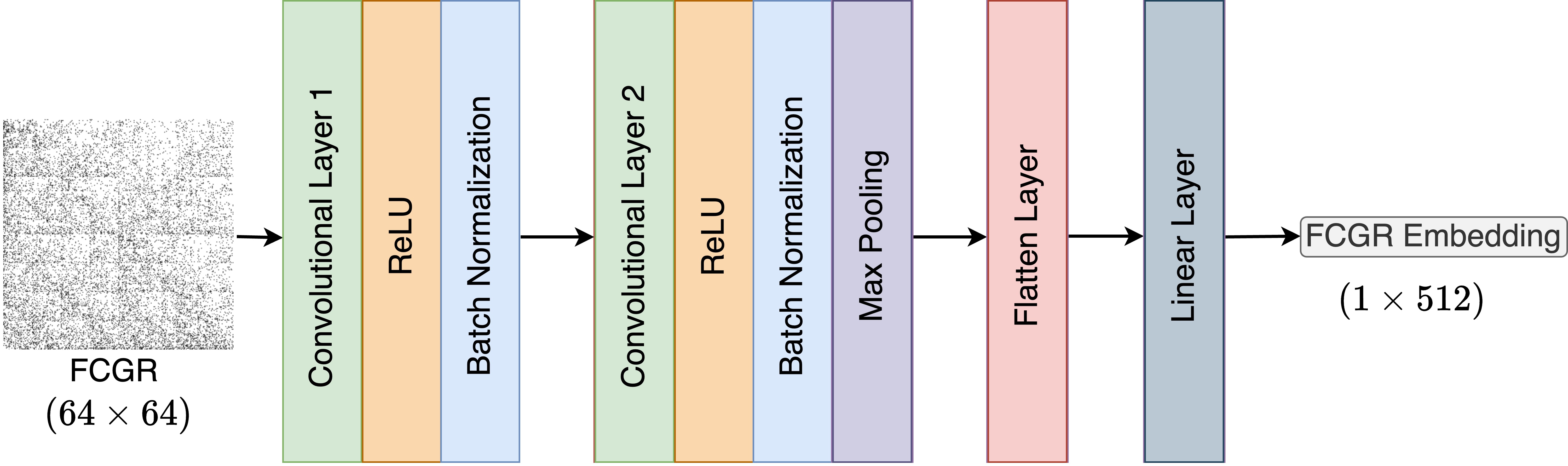}
\caption{\textbf{Architecture of backbone model designed for clustering FCGR images of DNA sequences.} The architecture of the backbone model comprises two convolutional layers, each with a kernel size of 7, stride of 2, and padding of 1. Following each convolutional layer, a Rectified Linear Unit (ReLU) is applied to introduce non-linearity, followed by a batch normalization layer to maintain numerical stability. Next, a max pooling layer with a kernel size of 2 efficiently reduces the spatial dimensions of the feature maps. A flattening layer to transform the multidimensional feature maps into a one-dimensional vector. This is followed by a linear layer, adjusting the output dimension to the desired configuration.} \label{fig:architecture}
\end{figure}

The augmented (mimic) DNA  sequence pairs of FCGRs ($X_{\tilde{s}_{2i-1}}, X_{\tilde{s}_{2i}}$) serve as inputs for training multiple independent instances of a backbone model, ICH, and CCH. Given that the genomic datasets we are working with are notably smaller in scale compared to those typically encountered in computer vision, we found that common architectures such as $ResNet34$ and $ResNet50$, which have demonstrated efficacy in various visual tasks, were not well-suited as backbone models for genomic datasets. Therefore, we opted for a simpler yet versatile architecture that is better suited for clustering FCGRs of DNA sequences. The backbone model architecture, as shown in Figure~\ref{fig:architecture}, is composed of a single convolutional block featuring two convolutional layers. Each convolutional layer employs a kernel size of 7, a stride of 2, and a padding of 1. Following each convolutional layer is a Rectified Linear Unit (ReLU) activation function and a batch normalization layer for data normalization prior to being passed to the subsequent layer. Subsequently, the output of the final batch normalization layer undergoes max pooling with a kernel size of 2 to downsample the data across its spatial dimension by selecting the maximum value within each $2\times2$ window. Lastly, to transform the multidimensional input into a one-dimensional embedding, a flattening layer is applied, followed by a linear layer configured to match the desired output dimension.

\subsection{Majority Voting Scheme}
The integration of ensemble learning, particularly through majority voting, has significantly improved the accuracy of genomic sequence classification, as demonstrated \textcolor{black}{by Mill{\'a}n et al.}~\cite{DeLUCS,iDeLUCS}. Majority voting, or hard voting, relies on the most frequent prediction across models, while soft voting considers the probability distributions of outcomes, often yielding higher precision. To optimize the performance of CGRclust, we employed five instances of the backbone model along with instance- and cluster-level contrastive heads. Each model copy was initialized randomly with distinct random seeds. Both soft and hard voting applied to CGRclust reduce variance due to random initialization and enhance model convergence thereby boosting the robustness and reliability of clustering predictions. Supplementary Material 2 discusses the impact of majority voting on clustering the Group 1 dataset. Although both voting methods enhanced CGRclust's performance, soft voting showed a slightly higher improvement. Consequently, we adopted soft voting as our default method. This approach integrates classifiers' certainty levels into the final prediction, thus yielding more reliable and potentially more accurate results.

\subsection{Experimental settings and implementation}
Throughout the training process, all CGRclust's hyperparameters remained constant and consistent across all tests, having been empirically chosen to achieve optimal performance. \textcolor{black}{We used the complexCGR library~\cite{complexCGR2024} to transform DNA sequences into their FCGR representations.} \textcolor{black}{We empirically chose $k=6$ for the resolution of FCGR after evaluating $k$ values ranging from 6 to 8. This selection offered an optimal trade-off between computational efficiency and accuracy.} Prior to input into the network, all FCGR \textcolor{black}{raw matrices} underwent normalization. This process involved first standardizing each FCGR matrix's value by the min-max normalization to scale the features to the range of  $[0, 1]$, thus mitigating the impact of sequence length on pixel intensity. Subsequently, the FCGR matrices were normalized by Z-score normalization to scale features so that they have the properties of a standard normal distribution with a mean of 0 and a standard deviation of 1. This normalization enhanced the stability and convergence of the model.

We utilized the Adam optimizer~\cite{Adam} with an initial learning rate set to $7\times10^{-5}$ and a weight decay of $10^{-4}$ to jointly optimize both contrastive heads and the backbone model. In our observations, the implementation of the scheduler did not yield significant improvements. Furthermore, the selection of batch size, empirically set at 512, is a critical factor during training. This importance stems from the batch-wise operation of the unsupervised learning process, which is essential for determining the output distribution. Inadequate batch sizes may fail to accurately represent the true data distribution, resulting in the dominance of the entropy term in the loss function and potentially leading to suboptimal solutions. The dimensionality of ICH was determined empirically to be 128, aiming to preserve discriminative information within the data. The dimensionality of CCH was determined by the target cluster number.

For benchmarking CGRclust's performance against state-of-the-art methods in DNA sequence clustering, we chose three recent alignment-free clustering methods noted for their effectiveness in clustering a variety of genomic datasets: DeLUCS~\cite{DeLUCS}, \textit{i}DeLUCS~\cite{iDeLUCS}, and MeShClust v3.0~\cite{meshclust}. For both DeLUCS and \textit{i}DeLUCS, we applied the default hyperparameters, and the accuracies presented in the results section are based on these settings. MeShClust v3.0, a density-based clustering tool, inherently does not allow the pre-definition of cluster numbers. Consequently, besides the automatic selection of identity thresholds—which often leads to a discrepancy between the expected and actual cluster counts—we tested several identity score thresholds to select an optimal value that resulted in the desired number of clusters for each dataset. The optimal threshold values for each of the thirteen real datasets tested are detailed in Supplementary Material 4. 

CGRclust's pipeline is fully implemented in Python 3.10, and the source code is publicly available in the GitHub repository \url{https://github.com/fatemehalipour/CGRclust}. All tests with CGRclust and DeLUCS were conducted on a node within the B\'eluga cluster at Compute Canada, which features dual Intel Gold 6148 Skylake CPUs @ 2.4 GHz, 186 GB RAM, and an NVIDIA Tesla V100 SXM2 GPU with 16 GB of memory. Following~\cite{iDeLUCS} authors' recommendation, \textit{i}DeLUCS was executed on Google Colab using an NVIDIA Tesla T4 GPU with 16 GB of memory.

\section{Results}\label{sec3}
\subsection{Qualitative performance of twin contrastive learning}
\begin{figure}[h]
\includegraphics[width=\textwidth]{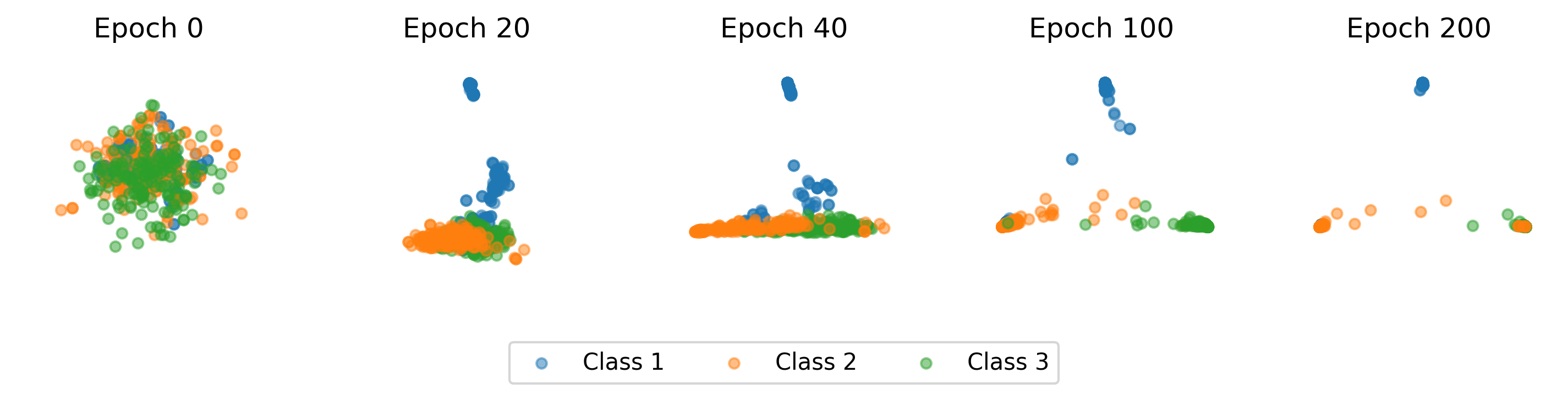}
\caption{\textbf{CGRclust's evolution of clustering 498 Cypriniformes mitochondrial DNA sequences into three distinct clusters in Test 1.} Each data point represents a DNA sequence, and its colour indicates its suborder label, and its position indicates the likelihood of assignment to different clusters (corners). A point at the center of the triangle has an equal probability of being assigned to any of the three clusters, while a point at a corner indicates a definitive association, with probability  1, to that specific corner/cluster. Note that any overlap of colors in the last epoch corresponds to instances of misclustering, where sequences have not been correctly assigned to the ground truth cluster. 
} \label{fig:learning-process}
\end{figure}

A qualitative analysis was first employed to assess the effectiveness of instance-level and cluster-level TCL, as implemented in CGRclust for clustering mtDNA sequences in Test 1. The dynamic learning process during the training phase is shown in Figure~\ref{fig:learning-process}, illustrating how the model develops discriminative representations and accurately determines cluster assignments. This progression is documented across epochs and displayed at five timestamps. In Figure~\ref{fig:learning-process}, the total number of clusters is established at three, corresponding to the points of a triangle, where each point signifies a taxonomic cluster. The placement of each point is derived from its three-dimensional probability vector, and different colors indicate the three ground truth taxonomic labels in Test 1. At the beginning, sequences are located at the triangle's center, reflecting an equal chance of being assigned to any of the three clusters. As training proceeds, the model increasingly assigns sequences to appropriate clusters, moving similar sequences closer to their respective vertex/cluster with greater probability. Notably, sequences that are assigned the same probability vectors will have their points overlap.

\subsection{Quantitative performance analysis and comparison with other methods}

In this section we analyze the performance of CGRclust and compare it with three other established clustering methods for DNA sequences, DeLUCS~\cite{DeLUCS}, $i$DeLUCS~\cite{iDeLUCS}, and MeShClust v3.0~\cite{meshclust} (with both manual and automatic selection of the identity score threshold). 
Note that the ground truth labels are used post-hoc and for evaluation purposes only, and they were not utilized during the clustering process.

Table~\ref{tab:comparison-Fish} \textcolor{black}{and Figure~S5.1} present a summary of the clustering accuracies for the Group 1 dataset described in Table~\ref{tab:merged-datasets} (Cypriniformes mtDNA) across Tests 1-4. The reader is referred to Supplementary Material 5 for the confidence intervals of the CGRclust clustering accuracies of all clustering tests. The accuracies of CGRclust were achieved using the default hyperparameters over 150 epochs. As Table~\ref{tab:comparison-Fish}, and Table~S5.1 in Supplementary Material 5 show, CGRclust consistently achieves comparable (within the confidence interval), or the highest accuracy across all four taxonomic levels. Specifically, CGRclust outperforms DeLUCS by  3.21\% to 12.95\% across different tests.  In contrast to the generally superior performance of CGRclust, \textit{i}DeLUCS shows competitive results in certain scenarios. Specifically, it achieves the highest accuracy among all methods at the suborder to family level  (92.06\%), comparable with CGRclust (within the confidence interval). This indicates that \textit{i}DeLUCS has particular strengths in clustering mtDNA datasets at some specific taxonomic levels. However, at other taxonomic levels, \textit{i}DeLUCS's performance generally is lower than both CGRclust and DeLUCS, suggesting that its clustering efficacy may vary depending on the nature and extent of sequence variation at a particular taxonomic level, and the characteristics of the dataset being analyzed. Lastly, CGRclust consistently outperforms both the manual and automated versions of MeShClust v3.0, by a large margin (up to 75.08\%). 
\begin{table}
\caption{\textbf{CGRclust performance of clustering the Group 1 dataset (Cypriniformes mtDNA) described in Table~\ref{tab:merged-datasets}.} CGRclust's accuracy is compared with DeLUCS, \textit{i}DeLUCS, and MeShClust v3.0 (with both a manual and the automatic selection of identity score threshold). Each row highlights the highest accuracy (within the confidence interval of CGRClust) in bold.}
\label{tab:comparison-Fish}
\centering
\setlength\tabcolsep{3pt}
\begin{tabular}{|c|c||c|c|c|c|c|}
\hline
\textbf{Test} & \textbf{Taxonomic} & \textbf{CGRclust} & \textbf{DeLUCS} & \textbf{\textit{i}DeLUCS} & \textbf{MeShClust-} & \textbf{MeShClust-} \\
 & \textbf{clustering}& & & &  \textbf{manual} & \textbf{auto}\\
\hline
1 & Order into suborder & \textbf{94.78\%} & 91.57\% & 64.05\% & 34.34\% & 33.94\% \\
\hline
2 & Suborder into family & \textbf{91.75\%} & 78.25\% & \textbf{92.06\%} & 20.16\% & 16.67\% \\
\hline
3 & Family into subfamily & \textbf{81.70\%} & 68.75\% & 61.61\% & 29.91\% & 12.5\%\\
\hline
4 & Subfamily into genus & \textbf{99.06\%} & 97.18\% & \textbf{99.53\%} & 59.15\% & 36.15\% \\
\hline
\end{tabular}
\footnotetext{The reader is referred to Table~S5.1 in Supplementary Material 5 for the confidence intervals of CGRclust clustering accuracies across Tests 1-4.}
\end{table}

Table~\ref{tab:comparison-Virus} \textcolor{black}{and Figure~S5.1} summarize the accuracies of clustering the five viral datasets in the Group 2 dataset described in Table~\ref{tab:merged-datasets} (viral whole genomes), across Tests 5-9. For the clustering of the astrovirus genomes (Tests 5 and 6), the clustering is at the family to genus level, while  for the dengue virus, HCV, and HIV-1 genomes the clustering is performed from the species to the virus subtype level. CGRclust consistently outperforms the other three clustering methods, demonstrating its robustness and accuracy in the context of virus mutagenesis and evolution. In Test 5, using an unbalanced astrovirus dataset, CGRclust surpasses DeLUCS and \textit{i}DeLUCS by 15.06\%, and outperforming MeShClust-manual by 25.25\%. The results demonstrate  CGRclust's  superior performance in challenging clustering tasks, e.g., characterized by dataset imbalance, a condition where other  methods  —DeLUCS, \textit{i}DeLUCS, and MeShClust v3.0— had a poor performance. In Test 6, which featured a cluster-balanced astrovirus dataset, the accuracy of both CGRclust and DeLUCS improved, while the accuracy of \textit{i}DeLUCS remained relatively unchanged. In the dengue virus genomes dataset (Test 7), CGRclust, along with DeLUCS and MeShClust-manual among the compared methods, achieved perfect accuracy (100\%) \textcolor{black}{without any errors}. For the HCV genome dataset (Test 8), CGRclust achieved an accuracy of 85.79\%, surpassing all compared methods by a margin of 1.16\% to 8.95\%. In the HIV-1 genomes dataset (Test 9), CGRclust achieves an accuracy that is 10.24\% higher than DeLUCS and significantly surpasses both \textit{i}DeLUCS and MeShClust-manual by 42.39\% and 48.1\%, respectively.

\begin{table}[h]
\caption{\textbf{CGRclust performance of clustering Group 2 dataset (viral whole genomes) described in Table~\ref{tab:merged-datasets}.} CGRclust's accuracy is compared with DeLUCS, \textit{i}DeLUCS, and MeShClust v3.0 (with both a manual and the automatic selection of identity score threshold). Each row highlights the highest accuracy (within the confidence interval of CGRClust) in bold. In Tests 5 and 6, the clustering task is at the family to the genus level, whereas Tests 7-9 involve clustering at the virus species to subtype level.}
\label{tab:comparison-Virus}
\centering
\setlength\tabcolsep{4pt}
\begin{tabular}{|c|c||c|c|c|c|c|}
\hline
\textbf{Test} & \textbf{Taxonomic} & \textbf{CGRclust} & \textbf{DeLUCS} & \textbf{\textit{i}DeLUCS} & \textbf{MeShClust-} & \textbf{MeShClust-} \\
 & \textbf{clustering}& & & &  \textbf{manual} & \textbf{auto}\\
\hline
5 & \begin{tabular}[t]{@{}c@{}}
Astroviridae-\\unbalanced
\end{tabular} & \textbf{84.94\%} & 69.88\% & 69.88\% & 59.69\% & 70.43\% \\
\hline
6 & \begin{tabular}[t]{@{}c@{}} 
Astroviridae- \\balanced
\end{tabular} & \textbf{88.84\%} & \textbf{88.84\%} & 69.97\% & 77.27\% & 76.72\% \\
\hline
\hline
7 & dengue virus & \textbf{100\%} & \textbf{100\%} & 96.99\% & \textbf{100\%} & 52.08\%\\
\hline
8 & hepatitis C virus & \textbf{85.79\%} & 84.63\% & 76.84\% & 81.05\% & 80.52\% \\
\hline
9 & \begin{tabular}[t]{@{}c@{}} 
human\\ immunodeficiency \\ virus 1
\end{tabular} & \textbf{81.77\%} & 71.53\% & 39.38\% & 32.77\% & 7.69\% \\
\hline
\end{tabular}
\footnotetext{The reader is referred to Table~S5.1 in Supplementary Material 5 for the confidence intervals of CGRclust clustering accuracies across Tests 5-9.}
\end{table}

Table~\ref{tab:comparison-idelucs} \textcolor{black}{and Figure~S5.1} display the clustering accuracies for Tests 10-13 in the Group 3 dataset (mtDNA of Insects, Protists, and Fungi) from the study~\cite{iDeLUCS}, detailed in Table~\ref{tab:merged-datasets}.  Due to the complexities and specific characteristics of datasets in the Group 3 dataset, we observed an enhancement in CGRclust performance when the hyperparameter \(\alpha\) was increased from its default  value of 0.7 to 0.8, along with a greater emphasis on the instance-level contrastive head. This modification is evidenced in the third and fourth columns of Table~\ref{tab:comparison-idelucs}, which display improvements in accuracy due to these adjustments. Generally, the change in the hyperparameter \(\alpha\) led to increased accuracy across this group of datasets, with the most notable improvement seen in the Protist dataset in Test 11, where accuracy rose by 23.28\%,  almost bridging the gap with DeLUCS and surpassing \textit{i}DeLUCS. However, in other datasets, this adjustment yielded minimal changes. This suggests that, in order to achieve optimal clustering outcomes, dataset-specific parameter optimization may be necessary to optimize different hyperparameters, including \(\alpha\).  Further details on hyperparameter adjustment of 
\(\alpha\) can be found in  Section 2.5 (Twin contrastive learning (TCL)).

In the comparison of clustering methods presented in Table~\ref{tab:comparison-idelucs} \textcolor{black}{and Figure~S5.1}, \textit{i}DeLUCS exhibits superior performance over other methods in the Insects mtDNA dataset of Test 10. However, both DeLUCS and CGRclust demonstrate higher accuracies in the other three tests. Specifically, in Test 11 (Protists mtDNA), the accuracies of DeLUCS and CGRclust are superior to \textit{i}DeLUCS by 8.10\% and 5.50\%, respectively. Furthermore, in Tests 12 and 13, both DeLUCS and CGRclust achieved higher accuracy in the Fungi classification at phylum and subphylum levels in comparison to \textit{i}DeLUCS and MeShClust v3.0. The manual and automatic versions of MeShClust generally display lower accuracies, with the automatic version particularly underperforming the manual selection of identity threshold in three out of four datasets. It is important to note that these datasets pose significant clustering challenges due to variations in within-cluster similarities and different sequence lengths, which complicate the clustering process. While CGRclust did not always secure the top clustering accuracy across these datasets compared to other methods, the adjusted version of CGRclust demonstrated comparable clustering performance in the Insects (Test 10) and Protists (Test 11) datasets, as well as the Fungi dataset at the subphylum level (Test 13). 
\begin{table}[h]
\caption{\textbf{CGRclust performance of clustering Group 3 dataset (mtDNA of Insects, Protists, and Fungi) described in Table~\ref{tab:merged-datasets}.} CGRclust (with and without an adjusted $\alpha$ hyperparameter) accuracy is compared with DeLUCS, \textit{i}DeLUCS, and MeShClust v3.0  (with both a manual, and the automatic selection of identity score threshold). Each row highlights the highest accuracy (within the confidence interval of CGRClust) in bold.}
\label{tab:comparison-idelucs}
\centering
\setlength\tabcolsep{1.5pt}
\begin{tabular}{|c|c||c|c|c|c|c|c|}
\hline
\textbf{Test} & \textbf{Taxonomic} & \textbf{CGRclust} & \textbf{CGRclust-} & \textbf{DeLUCS} & \textbf{\textit{i}DeLUCS} & \textbf{MeShClust-} & \textbf{MeShClust-} \\
 & \textbf{clustering} & & \textbf{adjusted $\alpha$} & &  & \textbf{manual} & \textbf{auto}\\
\hline
10 & Insects & 70.53\% & 73.56\% & 78.30\% & \textbf{83.82\%} & 47.50\% & 21.90\% \\
\hline
11 & Protists & 62.22\% & 85.50\% & \textbf{88.10\%} & 80.00\% & 71.85\% & 74.92\% \\
\hline
12 & \begin{tabular}[t]{@{}c@{}} Fungi\\(Phylum)\end{tabular} & 56.72\% & 56.87\% & \textbf{69.85\%} & 50.29\% & 50.74\% & 35.67\% \\
\hline
13 & \begin{tabular}[t]{@{}c@{}} Fungi\\(Subphylum)\end{tabular} & 97.10\% & \textbf{97.38\%} & \textbf{97.94\%} & 59.72\% & 75.14\% & 42.52\% \\
\hline
\end{tabular}
\footnotetext{The reader is referred to Table~S5.1 in Supplementary Material 5 for the confidence intervals of CGRclust clustering accuracies across Tests 10-13.}
\end{table}

Finally, for a direct comparison with MeShClust v3.0,  Table~\ref{tab:comparison-meshclust} \textcolor{black}{and Figure~S5.1} summarize the accuracies of clustering Group 4 dataset (the twelve synthetic datasets from~\cite{meshclust} and described in Table~\ref{tab:dataset-meshclust}), for all methods. In the Group 4 dataset, the terms ``Medium-'' and ``Long-'' in the dataset names indicate the sequence lengths. The numerical values ranging from 60 to 97 in the dataset names represent the identity score, a measure of sequence similarity. As this identity score increases, the sequences within a cluster become more similar, and this typically leads to enhanced performance of the clustering method. From the table, it is evident that CGRclust maintains a consistently high clustering accuracy, above 90\%, across both ``Medium'' and ``Long'' dataset categories. Although it does not always achieve the highest accuracy compared to the other methods, CGRclust's performance is relatively close to  DeLUCS and \textit{i}DeLUCS.

\begin{table}[h]
\caption{\textbf{CGRclust performance of clustering Group 4 dataset (the synthetic datasets from MeShClust v3.0~\cite{meshclust}), described in Table~\ref{tab:dataset-meshclust}.} CGRclust accuracy is compared with DeLUCS, \textit{i}DeLUCS, MeShClust v3.0.  The numerical values in the dataset names (in the range [60-97]) denote an identity score threshold signifying that every sequence within a cluster falls within this threshold distance from the cluster center. Each row highlights the highest accuracy (within the confidence interval of CGRClust) in bold.}
\label{tab:comparison-meshclust}
\centering
\begin{tabular}{|c|c||c|c|c|c|}
\hline
\textbf{Test} & \textbf{Dataset} & \textbf{CGRclust} & \textbf{DeLUCS} & \textbf{\textit{i}DeLUCS} & \textbf{MeShClust-auto} \\
\hline
14 & Medium-60 & 92.26\% & 94.97\% & 91.77\% & \textbf{99.7\%} \\
\hline
15 & Medium-70 & 93.39\% & 98.36\% & 94.40\% & \textbf{99.8\%} \\
\hline
16 & Medium-80 & 94.61\% & 99.42\% & 97.58\% & \textbf{99.8\%} \\
\hline
17 & Medium-90 & 95.23\% & 98.76\% & 97.60\% & \textbf{99.9\%} \\
\hline
18 & Medium-95 & 96.57\% & 99.73\% & 99.55\% & \textbf{100\%} \\
\hline
19 & Medium-97 & 95.51\% & 98.44\% & 98.57\% & \textbf{100\%} \\
\hline
20 & Long-60 & 93.31\% & 97.36\% & 94.13\% & \textbf{99.8\%} \\
\hline
21 & Long-70 & 92.82\% & \textbf{98.00\%} & 95.40\% & 93.41\% \\
\hline
22 & Long-80 & 96.29\% & 99.12\% & 97.03\% & \textbf{99.8\%} \\
\hline
23 & Long-90 & 94.08\% & 99.42\% & 99.13\% & \textbf{100\%} \\
\hline
24 & Long-95 & 94.20\% & 99.37\% & 99.67\% & \textbf{100\%} \\
\hline
25 & Long-97 & 94.83\% & 99.13\% & 99.11\% & \textbf{100\%} \\
\hline
\end{tabular}
\footnotetext{The reader is referred to Table~S5.1 in Supplementary Material 5 for the confidence intervals of CGRclust clustering accuracies across Tests 14-25.}
\end{table}  

\subsection{Summative Observations}
Overall, CGRclust exhibits versatility and robustness, consistently achieving high accuracy across twenty-five diverse datasets. CGRclust proved resilient to variations in dataset size, sequence length, and similarity, effectively handling the challenges posed by different genome types and taxonomic levels. Additionally, its performance in challenging scenarios, such as unbalanced datasets (e.g., Test 5), showcased its robust performance under different conditions.  Its consistent performance highlights its superior clustering capabilities and scalability compared to other established methods like DeLUCS, \textit{i}DeLUCS, and MeShClust v3.0. for DNA clustering.

The training duration for the twenty-five datasets varied, with the shortest being 413 seconds (almost 7 minutes) in Test 4, and the longest being 10,371 seconds (almost 3 hours) in Test 18, dependent on the sequence count. Notably, as CGRclust converts variable-length DNA sequences into fixed-size FCGRs, the training time remains relatively unaffected by sequence length. For detailed information regarding the total training time across all datasets, the reader is referred to Supplementary Material 6.

\section{Discussion}
This study explored the novel application of twin contrastive clustering of DNA sequences using Chaos Game Representation (CGR) to the field of bioinformatics, particularly to the unsupervised clustering of DNA sequences. The findings from this study provide a new perspective on the potential for unsupervised clustering methods, originally designed for computer vision, to achieve high accuracy in DNA classification/clustering tasks, traditionally dominated by supervised learning. 

Implementing this methodology required developing a robust algorithm capable of handling diverse genomic data types, ensuring consistent performance across different datasets, including fish mitochondrial genomes  (Cypriniformes order) at four taxonomic levels, as well as five different viral genomic datasets at genus or virus subtype levels. CGRclust achieved a high accuracy  even when used with an unbalanced dataset in Test 5 (the accuracy of CGRclust was 85\%, while the accuracies of the other methods were  15\% to 34\% lower), demonstrating its effectiveness in managing uneven data distributions. To ensure comprehensive evaluation and demonstrate the algorithm’s versatility, we expanded our dataset selection to include datasets previously analyzed by other studies (i.e., \textit{i}DeLUCS~\cite{iDeLUCS} and MeShClust v3.0~\cite{meshclust}). This inclusion allowed us to perform direct comparisons and validate the effectiveness of CGRclust across diverse genomic datasets. CGRclust successfully clustered all twenty-five tested datasets, which varied in length from 664 bp to approximately 100 kbp, covering a diverse range of cluster counts and sequence numbers. One of the primary challenges was optimizing the contrastive learning process to improve both the efficiency and accuracy of the clustering results. An effective pipeline that integrates data augmentation (generation of the \textit{mimic sequences}), feature extraction, and twin contrastive learning mechanisms successfully addressed this issue. It is important to note that, although this study focused on DNA sequences in the clustering experiments, CGRclust could also be applied to RNA analysis. This is due to the fact that both DNA and RNA are sequences made up of four ``letters,"   that can  each act as the label of one of the four corners of a CGR square.

The applicability of our method has been primarily evaluated using the datasets mentioned, but further extensive validation across a wider range of DNA clustering tasks is necessary. This includes testing on DNA sequences longer than 100 kb, with a higher number of genome sequences per cluster, and a greater number of clusters, to confirm its general applicability. Beyond taxonomic clustering, this method could also be explored in other contexts such as  exploring the impact of extreme environments on  genomic signatures, and virus-host genomic signature similarity.

Additionally, while CGRclust is more time-efficient compared to alignment-based methods and comparable to other clustering methods evaluated, it can still be time-consuming, especially when applied to large datasets. This limitation, which comes from the substantial batch sizes required for effective contrastive learning, could limit CGRclust's practicality in settings where rapid processing of genomic data is required. \textcolor{black}{For the purpose of rapidly estimating evolutionary distances for closely related sequences without relying on labelled data, other tools such as andi~\cite{10.1093/bioinformatics/btu815} and phylonium~\cite{klotzl2020phylonium} exist. As detailed in Supplementary Material~7, our experiments confirmed that phylonium performs efficiently on datasets used in Tests 1, 2, 3, and 4 (mtDNA of Cypriniformes), and Test 9 (HIV-1 genomes), generating evolutionary distance matrices in under a minute. However, for the remaining twenty datasets, characterized by more heterogeneous sequences, phylonium aborted the task and generated matrices with NaN values. This demonstrates that CGRclust is applicable to a wider range of datasets than phylonium, as it effectively clusters datasets containing dissimilar and non-alignable sequences that cannot be classified by tools optimized for closely related sequences.  Ultimately, the determination of the optimal clustering tool has to be guided by the specifics of the application.}

Another limitation of CGRClust is finding a set of hyperparameters that is universally effective across different types of tests, which has proven to be challenging and may indeed be impossible given the diversity in genomic data and clustering objectives. In other words, each type of dataset may require individual finetuning of the model's hyperparameters in order to achieve optimal accuracy, and this can significantly increase the complexity and duration of the initial set-up.

In light of these limitations, future work should focus on optimizing the computational efficiency of the method, exploring its scalability across diverse genomic datasets, and developing adaptive hyperparameter tuning mechanisms that can respond dynamically to the characteristics of the data being processed.

\section{Conclusions}
This study introduces CGRclust, a novel twin contrastive clustering algorithm for the taxonomic clustering of unlabelled  DNA sequences. CGRclust utilizes unsupervised machine learning to 
identify relevant and discriminative patterns in unlabelled, primary DNA sequence data, without relying on homology, sequence alignment, or any biological and taxonomic labelling. CGRclust achieves high clustering accuracies by combining the visual Chaos Game Representation of DNA sequences, with  recent advancements in unsupervised learning for computer vision, namely twin contrastive learning and convolutional neural networks. It successfully clusters different datasets including full mitochondrial DNA genomes from fish, fungi, protists, and viral whole genomes across different taxonomic levels 
from phyla to intraspecific subtypes. Remarkably, CGRclust obtained high accuracy when encountering cluster imbalance in a dataset, showcasing its robustness with uneven data distributions. CGRclust achieves higher or comparable clustering accuracies compared with state-of-the-art existing unsupervised machine learning clustering methods, across all datasets tested. Notably, in 11 out of 13 real datasets, CGRclust achieved accuracy greater than 80\%. In comparison, the DeLUCS algorithm surpassed this accuracy threshold in 7 out of 13 tests, \textit{i}DeLUCS in only 5 tests, and MeShClust v3.0 only once. This demonstrates that CGRclust's performance is more consistently reliable than other methods. In particular, CGRclust  performed especially well on viral datasets, where it consistently achieved the highest accuracies.

\section{List of abbreviations}

\textbf{CCH:} Cluster-level Contrastive Head\\
\textbf{CGR:} Chaos Game Representation\\
\textbf{CNN:} Convolutional Neural Network\\
\textbf{FCGR:} Frequency Chaos Game Representation\\
\textbf{HCV:} hepatitis C virus\\
\textbf{HIV:} human immunodeficiency virus\\
\textbf{ICH:} Instance-level Contrastive Head\\
\textbf{KNN:} K-Nearest Neighbor\\
\textbf{mtDNA:} Mitochondrial DNA\\
\textbf{NCBI:} National Center for Biotechnology Information\\
\textbf{ReLU:} Rectified Linear Unit\\
\textbf{SVM:} Support Vector Machine\\
\textbf{TCL:} Twin Contrastive Learning\\

\section*{Declarations}

\begin{itemize}
\item \textbf{Ethics approval and consent to participate}\\
      Not applicable.\\
\item \textbf{Consent for publication}\\
      Not applicable.\\
\item \textbf{Availability of data and material}\\
      The datasets generated and/or analyzed during the current study are all available in public repositories, and the links can be found in section 2.1 (Datasets) or associated literature. The CGRclust method developed for this study, along with all datasets used are available at \url{https://github.com/fatemehalipour/CGRclust}.\\ 
\item \textbf{Competing interests}\\
      The authors declare no competing interests.\\
\item \textbf{Funding}\\
      The authors declare financial support was received for the research, authorship, and/or publication of this article. This work was supported by Natural Science and Engineering Research Council of Canada Grants RGPIN-2023-05256 to K.A.H. and RGPIN-2023-03663 to L.K. This research was enabled in part by support provided by Compute Canada RPP (Research Platforms Portals), https://www.computecanada.ca/, Grant 616 to K.A.H. and L.K. The funders had no role in the preparation of the manuscript.\\
\item \textbf{Authors' contributions}\\
      F.A., and L.K. conceived the study and wrote the manuscript. F.A. designed and performed the experiments. F.A., L.K., and K.A.H. conducted the data analysis and edited the manuscript, with K.A.H. contributing biological expertise. All authors read and approved the final manuscript.\\
\item \textbf{Acknowledgements}\\
We thank Dr. R. Greg Thorn for his guidance on fungi taxonomy, Matheus Sanita Lima for guidance on protist taxonomy, Joseph Butler for proofreading the manuscript, and Pablo Mill{\'a}n Arias for his assistance with experiments with \textit{i}DeLUCS.
\end{itemize}

\bmhead{Supplementary information}
\begin{itemize}
\item \href{https://github.com/fatemehalipour/CGRclust/blob/main/supplementary/S1.pdf}{Supplementary Material 1: Chaos Game Representation}
\item \href{https://github.com/fatemehalipour/CGRclust/blob/main/supplementary/S2.pdf}{Supplementary Material 2: CGRclust Methodological Optimization}
\item \href{https://github.com/fatemehalipour/CGRclust/blob/main/supplementary/S3.pdf}{Supplementary Material 3: Twin Contrastive Learning}
\item \href{https://github.com/fatemehalipour/CGRclust/blob/main/supplementary/S4.pdf}{Supplementary Material 4: Optimal Threshold Values Across Datasets for MeShClust v3.0}
\item \href{https://github.com/fatemehalipour/CGRclust/blob/main/supplementary/S5.pdf}{Supplementary Material 5: CGRclust Comparative Clustering Accuracies}
\item \href{https://github.com/fatemehalipour/CGRclust/blob/main/supplementary/S6.pdf}{Supplementary Material 6: CGRclust Training Times Across Different Tests}
\item \href{https://github.com/fatemehalipour/CGRclust/blob/main/supplementary/S7.pdf}{Supplementary Material 7: Comparative Evaluation of Phylonium}
\end{itemize}

\bibliography{main}

\end{document}